\documentclass[12pt]{article}   
 \usepackage{psfig,times,fancyheadings}
        
\begin{document}
\thispagestyle{empty} 


 \lhead[\fancyplain{}{\sl }]{\fancyplain{}{\sl }}
 \rhead[\fancyplain{}{\sl }]{\fancyplain{}{\sl }}


\newcommand{\nc}{\newcommand}

\nc{\qI}[1]{\section{ {#1} }}
\nc{\qA}[1]{\subsection{ {#1} }}
\nc{\qun}[1]{\subsubsection{ {#1} }}
\nc{\qa}[1]{\paragraph{ {#1} }}

\nc{\qfoot}[1]{\footnote{ {#1} }}
\def\qL{\hfill \break}
\def\qpar{\vskip 2mm plus 0.2mm minus 0.2mm}
\def\tvi{\vrule height 12pt depth 5pt width 0pt}

\def\qparr{ \vskip 1.0mm plus 0.2mm minus 0.2mm \hangindent=10mm
\hangafter=1}

\def\qdec#1{\par {\leftskip=2cm {#1} \par}}

\def\qdpt{\partial_t}
\def\qdpx{\partial_x}
\def\qddpt{\partial^{2}_{t^2}}
\def\qddpx{\partial^{2}_{x^2}}
\def\qn#1{\eqno \hbox{(#1)}}
\def\qds{\displaystyle}
\def\qal{\sqrt{1+\alpha ^2}}
\def\qw{\widetilde}


\null
\vskip 1cm
{\bf \sl \large To appear in: The European Physical Journal B} 
\vskip 2cm

\centerline{\bf \LARGE Determining bottom price-levels}
\vskip 0.5cm
 \centerline{\bf \LARGE after a speculative peak}

\vskip 1cm
\centerline{\bf B.M. Roehner $ ^* $ }
\centerline{\bf L.P.T.H.E. \quad University Paris 7 }

\vskip 2cm

{\bf Abstract}\quad  During a stock market peak the price of a given 
stock ($ i $) jumps from an initial level $ p_1(i) $ to a peak level
$ p_2(i) $ before falling back to a bottom level $ p_3(i) $. The
ratios $ A(i) = p_2(i)/p_1(i) $ and $ B(i)= p_3(i)/p_1(i) $ are referred to
as the peak- and bottom-amplitude respectively. The paper
shows that for a sample of stocks there
is a linear relationship between $ A(i) $ and $ B(i) $ of the
form: $ B=0.4A+b $. In words, this means that the higher the price of
a stock climbs during a bull market the better it resists during the
subsequent bear market. That rule, which we call the resilience pattern,
also applies to other speculative markets.
It provides a useful guiding line for Monte Carlo simulations.

\vskip 1cm
  \centerline{\bf  5 June  2000 }

\vskip 1cm
{\bf PACS.}\ 64.60 Equilibrium properties near critical points - 
 87.23Ge Dynamics of social systems

\vskip 2cm
* Postal address: LPTHE, University Paris 7, 2 place Jussieu, 75005 Paris, France
\qL
\phantom{* }E-mail: ROEHNER@LPTHE.JUSSIEU.FR
\qL
\phantom{* }FAX: 33 1 44 27 79 90

\vfill \eject

\qI{Introduction}

Traffic jams are fairly unpredictable because they depend upon a large
number of factors, e.g. timing in the traffic, weather conditions,
highway maintenance, automobile accidents, some of which are completely
random. However once begun, traffic jams display fairly recurrent patterns
as to average duration, behavior of the drivers, and so on. This
traffic jam parallel has been introduced by Charles Tilly [12] in
the context of historical sociology in order to explain why revolutions
can hardly be predicted. It also applies to the occurrence of speculative
price peaks: the downturn of price peaks can hardly
be predicted because they depend upon a number of (possibly)
exogenous factors%
\qfoot{For instance it has been argued (Business Line 8 May 2000) that 
the spurt in the NASDAQ composite index that occurred in December 1999 and
January 2000 was fueled by a Y2K-motivated injection of money into the
banking system. Needless to say, no model will ever be able to take 
into account such circumstantial factors. However, the present paper
suggests that when a market rallies or plummets 
there are some structural invariants 
which are independent of triggering factors.}%
. 
However, as is the case for traffic jams, once a price peak is under 
way it obeys some definite rules; accordingly its outcome can to some
extent be predicted at the level of individual companies. 
\qL
More precisely
we focus our attention on the relationship between the prices at the
beginning of the peak, at the peak and at the end of the peak. We 
denote by $ p_1 $ the price level at the start of the peak, by $ p_2 $
the price at the peak and by $ p_3 $ the bottom price at the end
of the falling price path; we further introduce the peak amplitude
$ A = p_2/p_1 $ and the bottom amplitude $ B = p_3/p_1 $. 
$ A $ and $ B $ can be defined for any company in the market; for instance
on the NASDAQ where there are currently more than 5,000 companies listed 
$ A $ and $ B $ can be seen as variables for which there
are several thousand realizations. In the next section we show that 
$ A $ and $ B $ are closely correlated; with a correlation of the order
of 0.75 the following regression holds:
 $$ B = aA + b \qn{1} $$

where $ a $ is usually of the order of 0.4. The fact that $ a $ is positive
means that the higher the peak amplitude, the larger the bottom amplitude;
in other words the higher the price of a stock climbs during the
rising phase (bull market) the better it resists during the falling
phase (bear market). The regularity summarized by equation (1) will
be referred to as the resilience pattern.
\qL
The paper proceeds as follows. In the second section the statistical
methodology is explained, then the resilience pattern is established
and illustrated through several case studies: we consider three
stock market peaks, one price bubble for postage stamps and two 
real estate bubbles. In the conclusion we discuss the possible 
implications of the resilience pattern.

\qI{The resilience pattern}
\qA{Methodology}
Let us consider a typical price peak such as the one in Fig.1. 
While it is easy to identify the moment $ t_2 $ when the price reaches
its peak level $ p_2 $, the determination of the initial moment $ t_1 $
(corresponding to $ p_1 $) and the end moment $ t_3 $ (corresponding
to $ p_3 $) is not so easy. Fortunately, as will be seen subsequently,
the relationship (1) does not depend upon the choice of $ t_1 $ or
$ t_2 $ in a critical way. 
\qL
A peak will be delimited in two steps (i) At the global level of the
whole market the price peak is identified by using a broad annual
index; in that way one already gets a rough definition of
$ t_1, t_2, t_3 $ (ii) At the level of individual companies $ t_1 $
will be selected as the first year for which the 
annual price change is positive,
$ t_2 $ as the peak year and $ t_3 $ as the last year for which the 
annual price change is negative. Let us illustrate the procedure on an
example (Fig.1). As one knows there was a short bull market on the
NYSE after World War I which culminated in 1920; 
thus, for the great majority of the stocks the first positive price change
was 1921-1922, which leads us to $ t_1=1921 $. At the end of 1929 the
downturn occurred very abruptly which means that for almost all stocks
$ t_2=1929 $. By and large the market bottomed out in 1931; yet for some
stocks such as  for instance Consolidated Edison
the fall continued until 1935; in that case one would take $ t_3=1935 $. 
\qL
Once the limits of the peak have been 
determined for each individual stock,
the ratios $ p_2/p_1 $ and $ p_3/p_1 $ are computed for
the deflated prices; then the correlation and regressions are 
carried out for the sample of stocks under consideration. We now
apply this procedure to several case studies.

\qA{Stock markets}
As far as stock markets are concerned there is often a tendency to
over-emphasize the importance of crashes, by which we understand
a rapid fall occurring within one or two weeks, at the expense of the
long and steady declines that (in some cases) follow the crash. Crashes 
are impressive because of their suddenness, but it is not obvious 
and probably not even true, that crashes have a determining influence
on the medium-term (i.e. yearly) evolution of markets. A slide that 
continues for over two or three years may have more significance for 
the stock market and for the rest of the economy 
than the crash itself.
A case in point is the parallel between the crashes of
October 1929 and October 1987 on the NYSE (Fig.2). The price paths were
very similar during the crashes and the six subsequent months;
but, as one knows, the ultimate outcomes were very different. 
This observation
suggests that 
steady slides (which are the topic of this paper) and abrupt crashes 
are two different phenomena. 
\qL
Applying the procedure delineated above, we obtain the results given
in Table 1. Note that the downturn in Paris occurred in February 1929
and cannot therefore be considered as a consequence of the Wall Street
crash; incidentally, in Germany the downturn took place in June 1928
that is to say more than one year before the downturn on the NYSE.
The analysis for the NYSE is based on the behavior of 85 individual
stocks. The correlation is 0.87 for 
equation (1) and the distribution of sample points
in the $ (A,B) $ plane is shown in Fig.3; it displays the range
of both amplitudes and permits to verify that there is no non-linear
effect. 
\qL
Since this is the largest sample considered in Table 1, it can
can be of interest to take a closer look at the statistical distribution
of the amplitudes: $ A $ and $ B $ have an average of 5.6 and 2.0 
respectively and the standard deviations are 5.0 and 2.3. Moreover it 
turns out that both $ A $ and $ B $ are distributed according to
a log-normal density. This could have been expected; indeed, 
stock prices  follow a log-normal law, 
at least in first approximation
that is to say for time-samples of moderate size
and time intervals larger than one day, 
and one knows that the ratio of two log-normal random variables
is also a log-normal random variable. 
\qpar

The analysis of the Paris stock market is not based on individual stocks
but on a set of indexes corresponding to 19 different economic sectors,
e.g. banks, coal mines, railroads, electricity, chemicals, etc. Some indexes
comprise more individual stocks than others, for instance the bank index
comprises 20 banks while the electricity index has only 11; on average there
are about 14 companies per sector. The results for the $ B $ versus
$ A $ regression are given in Table 1: the $ a $ estimate is fairly
close to the one obtained for the NYSE. The analysis of the 1989 
peak on the Tokyo market is also based on indexes corresponding to 
different economic sectors. 
With as many as 26 different sectors the classification given in the
Japan Statistical Yearbook is even more detailed than the previous one.
It is not obvious whether that peak began in 1985 or in 1980.
It is true that the increase between 1980 and 1985 was not monotonic,
but one can argue with good reason that these fluctuations were
rather circumstantial. It is reassuring to observe that by taking
$ t_1=1980 $ one is lead to estimates which are fairly similar to those
obtained for $ t_1=1985 $; the fact that the correlation is higher 
in the first case
in fact suggests that $ t_1=1980 $ is the most ``natural'' starting point.
\qpar

On the basis of the resilience pattern it could seem that one can
invest in high-growth companies without much risk. This is not completely
true however for that rule only concerns the medium-term 
behavior in the
vicinity of a given peak; it does not guarantee that in the long-run the
price of a stock which has experienced a huge peak will continue to increase.
A spectacular counter-example is shown in Fig.4. Not only did the
price of Columbia Gas System never again reach the level it had
attained in 1929 but it remained far below. 

\qA{Other speculative markets}
The consideration of other speculative markets relies on the implicit
assumption that the {\it basic} mechanisms of speculation are similar
for any speculative market. In this paragraph we examine the cases
of postage stamp and real
estate markets. These markets are particularly suited
for this kind of investigation because (i) they have large price 
peaks and (ii) they comprise a large number of different items which
will play the same role as individual stocks or economic sectors in our
previous study. 
\qpar

During Word War II there was a postage stamp bubble in France which
was triggered by the fact that inflation was high (the consumer price
index increased from 130 in 1941 to 350 in 1945) while at the same time,
due to war restrictions there was only a small outlet for consumption.
Thus almost all stamps experienced a price peak (usually culminating in
1944) with their real (i.e. deflated) price multiplied by a factor
of 3 or 4. The present investigation focuses on 19th century French stamps.
Table 1 shows that the correlation between peak and bottom amplitude
although still significant is smaller than in previous cases. This can
possibly be attributed to the fact that the prices given in 
stamp catalogs are estimates made by stamp experts and traders rather than
real market prices. Estimates are particularly difficult to make for stamps
which are not traded very often; such is the case for the stamps which
are particularly rare and expensive. Now (see in this respect 
[11] 
) the expensive stamps are also those for which the
peak amplitude $ A $ is largest and any bias for large $ A $ sample
points will notably affect the correlation and the slope of the regression
line. 
\qpar

Our next example concerns the real estate market in Paris. Between
1985 and 1995 there was a price peak which resulted in a doubling, and
in some areas a three fold increase, of apartment prices. First, we consider
the prices in 20 different districts (``arrondissements'')
of downtown Paris (i.e. the so-called ``Paris intra-muros'' 
area). 
The regression leads to values
for the correlation and for the slope of the regression line which
are similar to those obtained in the case of stock markets. 
As shown in Fig.5 the sample points in the $ (A,B) $ plane are evenly
distributed along the regression line and do not display any obvious
non-linear effect. 
\qL
The second result concerns the same price peak but for apartments 
according to their size (from one- to five-room). Not surprisingly, due
to the small number of sample points the error margin is fairly large;
note that the value $ a =0.4 $ obtained in previous cases is within the
error bars. 
\qpar

Up to that point all our results were consistent with the resilience
pattern, but the following case lead to an unexpected result. It 
concerns the real estate
bubble which occurred in Britain in the 1980s; from
the area of London where it started, it spread progressively northward
to the rest of the country. Price results can be analyzed at the level
of each of the 11 regions composing Britain. The result came as a surprise;
in this case
the amplitudes are negatively correlated and the correlation
is fairly low. In order to see if the negative correlation is found elsewhere
it would be of great interest to perform a similar test for other countries;
unfortunately, in France reliable regional housing prices are only available
since 1995.

\begin{table}[t]

\centerline{\bf Table 1 \ The resilience pattern: relationship between} 
\centerline{\bf peak amplitude $ (A) $ and bottom amplitude 
                   $ (B) $: $ B=aA+b $}

\vskip 5mm
\hrule
\vskip 0.5mm
\hrule
\vskip 2mm

$$ \matrix{
\tvi & \hbox{Market} & \hbox{Peak} & \hbox{Number} & a & b &
\hbox{Correlation} \cr
\tvi & \hbox{} & \hbox{} & \hbox{of items} &  &  & \hbox{} \cr
\noalign{\hrule}
 \tvi & \hbox{Stocks} \hfill & \hbox{} & \hbox{} &  &  & \hbox{} \cr
1 & \hbox{NYSE} \hfill& 1929\hbox{ Oct} & 85 & 0.40 \pm 0.05 & -0.27 \pm 0.24 & 
 0.87 \cr
2 & \hbox{Paris}\hfill & 1929\hbox{ Feb} & 19 & 0.44 \pm 0.17 & -0.03 \pm 0.29 & 
 0.76 \cr
3a & \hbox{Tokyo}\hfill & 1989\hbox{ Oct} & 26 & 0.39 \pm 0.11 & 0.27 \pm 0.10 & 
 0.80 \cr
3b & \hbox{Tokyo}\hfill & 1989\hbox{ Oct} & 26 & 0.40 \pm 0.09 & 0.35 \pm 0.13 & 
 0.87\cr
 & \hbox{} \hfill & \hbox{} & \hbox{} &  &  & \hbox{} \cr
 & \hbox{Stamps} \hfill & \hbox{} & \hbox{} &  &  & \hbox{} \cr
4 & \hbox{France}\hfill & 1944 \hbox{} & 56 & 0.10 \pm 0.08 & 0.77 \pm 0.10 & 
 0.33 \cr
 & \hbox{} \hfill & \hbox{} & \hbox{} &  &  & \hbox{} \cr
 & \hbox{Real estate} \hfill & \hbox{} & \hbox{} &  &  & \hbox{} \cr
5 & \hbox{Paris}\hfill & 1990 \hbox{} & 20 & 0.43 \pm 0.16 & 0.73 \pm 0.04 & 
 0.79 \cr
6 & \hbox{Paris}\hfill & 1990 \hbox{} & 5 & 0.14 \pm 0.26 & 1.1 \pm 0.1 & 
 0.52 \cr
7 & \hbox{Britain}\hfill & 1989 \hbox{} &11 & -0.10 \pm 0.17 & 1.4 \pm 0.05 & 
 -0.38 \cr
 & \hbox{} \hfill & \hbox{} & \hbox{} &  &  & \hbox{} \cr
 & \hbox{Average}\hfill & \hbox{(except 7)} &  & 0.32 &  &  0.68 \cr
} $$

\hrule
\vskip 0.5mm
\hrule
\vskip 5mm

{\small Notes: All prices used in the regression are real 
(i.e. deflated) prices. 
Case (3a) refers to $ t_1=1985 $, while case (3b) refers to $ t_1=1980 $.
For some reason yet to be understood the housing bubble in Britain does 
not follow the resilience pattern (negative correlation). 
\qL
{\sl Sources: 1: [1]; 2: Annuaire Statistique de la
France, R\'esum\'e R\'etrospectif (1966, p.541); 3a,b: Japan
Statistical Yearbook (various years); 4: [5]; 5,6: Conseil par des notaires
(23 Dec. 1991) and Chambre des Notaires; 7: Halifax index.}}

\end{table}

\qI{Conclusion}

About 60 percent of the 222 sample points in Table 1 concerned stock
markets. Thus, the evidence supporting the resilience pattern 
for stock markets was particularly strong. Yet, it was important also
to show that the resilience effect is {\it not} confined to stock
markets for it suggests that a possible theoretical framework should
apply to other speculative markets as well. Let us now briefly
discuss the significance and possible implications of the present finding.
\qL
In the late 1990s econophysicists along with some economists ([3])
devoted great attention to the 
statistical analysis of stock market indexes, the overall objective
being the identification of possible scaling laws. Yet, indexes do not
give great insight into the internal mechanisms of stock markets. Such
an understanding can only be gained by opening the ``black box'' 
and studying the interactions that take place between individual stocks.
This idea has recently gained more acceptance as shown by a number
of innovative papers going into that direction; e.g. 
[3, 
4, 
6 
]. 
\qL
Finally, let us briefly consider the next step, namely the construction
of a theoretical framework. Obviously any model is 
(and has to be) a schematization of the real world; therefore, 
constructing a ``realistic'' model cannot be a viable and suitable 
objective; models need more precise ``targets'' and ``guiding lights''.
In a number of recent empirical studies we have tried to define
such targets: the sharp peak - flat trough pattern 
[9, 10], 
the price multiplier effect
[7, 9], 
the relationship between stock market crashes and increases in interest
rate spread
[8] 
define quantitative patterns which provide useful guiding lights
for the construction of a theoretical framework. 
\qL
On the theoretical side some promising advances have been made recently
which can possible provide an adequate framework for the description of the 
internal machinery of stock markets. For instance one would not be
surprised to see percolation  (see in this respect
[2]) 
play a role in the spread of a bubble; 
after all, a speculative outburst 
can be seen as propagating from high-growth stocks
to low-growth stocks in the  same way as a technical
innovation progressively
gains acceptance in an economy.
\qpar

{\bf Acknowledgments}\quad I would like to express my gratitude to the
statistical experts of the Halifax Company (U.K.) 
and the Chambre des Notaires
(Paris) for their kind assistance.

\vfill \eject

\centerline{\bf \Large References}

\vskip 1cm

\qparr
(1) COMMON STOCK price histories 1910-1986 and logarithmic supplement, 1988:
WIT Financial Publishers. Anchorage.

\qparr
(2) GOLDENBERG (J.), LIBAI (B.), SOLOMON (S.), JAN (N.), STAUFFER (D.) 2000:
 Marketing percolation. Physica A (to appear). Cond-mat/0005426.

\qparr
(3) LUX (T.) 1996: The stable Paretian hypothesis and the frequency of
large returns: an examination of major German stocks. 
Applied Financial Economics 6,463-475.

\qparr
(4) MANTEGNA (R.N.) 1999: Hierarchical structure in financial markets. 
European Physical Journal B 11,1,193-197.

\qparr
(5) MASSACRIER (A.) 1978: {\sl Prix des timbres-poste classiques de 1904 \`a 1975.}
A. Maury. Paris.

\qparr
(6) PLEROU (V.), GOPIKRISHNAN (P.), AMARAL (L.A.N.), MEYER (M.), STANLEY (H.E.)
1999: Scaling of the distribution of price fluctuations of individual
companies. Physical Review E 60,6,6519-6529.

\qparr
(7) ROEHNER (B.M.) 2000: Speculative trading: the price multiplier effect.
The European Physical Journal B 14,395-399.

\qparr
(8) ROEHNER (B.M.) 2000: Identifying the bottom line after a stock market
crash. International Journal of Modern Physics 11,1,91-100.

\qparr
(9) ROEHNER (B.M.) 2001: {\sl Hidden collective factors in speculative trading.}
Springer-Verlag. Berlin (to appear).

\qparr
(10) ROEHNER (B.M.), SORNETTE (D.) 1998: The sharp peak - flat trough pattern
and critical speculation. The European Physical Journal B 4,387-389.

\qparr
(11) ROEHNER (B.M.), SORNETTE (D.) 1999: Analysis of the phenomenon of
speculative trading in one of its basic manifestations: postage stamp 
bubbles. International Physical Journal B (to appear).

\qparr
(12) TILLY (C.) 1993: {\sl European revolutions 1492-1992.} 
Blackwell. Oxford.

\vfill \eject

 \def\qpage{\vfill \eject \ \vfill \eject}

\def\qv{\vskip 0.1mm plus 0.05mm minus 0.05mm}
\def\qhu{\hskip 0.6mm}
\def\qhv{\hskip 3mm}
\def\qhw{\hskip 1.5mm}

\baselineskip=11pt

\def\qleg#1#2#3{ \noindent
{\bf \small #1\qhw}{\small #2\qhw}{\it \small #3}\qv }

\qleg{Fig.1\qhv Course of the price for three stocks versus Dow Jones
Index.}
{The prices are yearly highs. During the bull market 1921-1929, the 
prices of Coca Cola and Columbia Gas (an utility company) have 
increased faster and more than the DJ average, while the price of
Burlington Northern (a railroad company)
has increased slower and less than the DJ. 
The circles and squares show the peaks and troughs respectively.
The resilience pattern reveals itself in the fact that when a peak is above
(or below) the DJ average the same situation prevails for the 
trough.}
{Sources: Common stock price histories (log supplement); Dow Jones
Investor's handbook (1972).}

\qleg{Fig.2\qhv Parallel between the crashes of 1929 and 1987.}
{Monthly prices. During the 8 months before and after the crash
the behavior of the Dow Jones Index was very much the same in
both cases; as one knows the subsequent evolution was very different
however. This suggests that crashes and long-lasting slides are
two distinct (and not necessarily related) phenomena.}
{Sources: The Dow Jones averages 1885-1970; OECD main economic 
indicators (1969-1988).}

\qleg{Fig.3\qhv NYSE, 1921-1932 peak:  distribution of 85 sample
points in the $ (A,B) $ plane.} 
{Each triangle corresponds to a stock;
the prices are yearly highs. Dotted line: linear regression (correlation
is 0.87).}
{Source: Common stock price histories (log supplement).}

\qleg{Fig.4\qhv Long-term behavior of the price of Columbia Gas System.}
{The prices are yearly highs.
The price of that utility stock increased tremendously during the
1921-1929 bull market; then, from 1932 to 1936 it fluctuated around
200 before falling to about 30; it never really recovered in the
second half of the 20th century. The chart does not cover the
1990s because there was a change in the name of the company; however
it can be noted that during the 1990-2000 bull market the Dow Jones
Utilility index increased 1.7 times less than the Standard and Poor's 500;
this would give a stock price of about 200 in year 2000, still well
below the 1929 high.}
{Source: Common stock price histories (log supplement).}

\qleg{Fig.5\qhv Paris, 1985-1995 real estate peak: distribution of 20 
sample points in the $ (A,B) $ plane.} 
{The numbers correspond to the 20 arrondissements.}
{Source: Chambre des Notaires.}

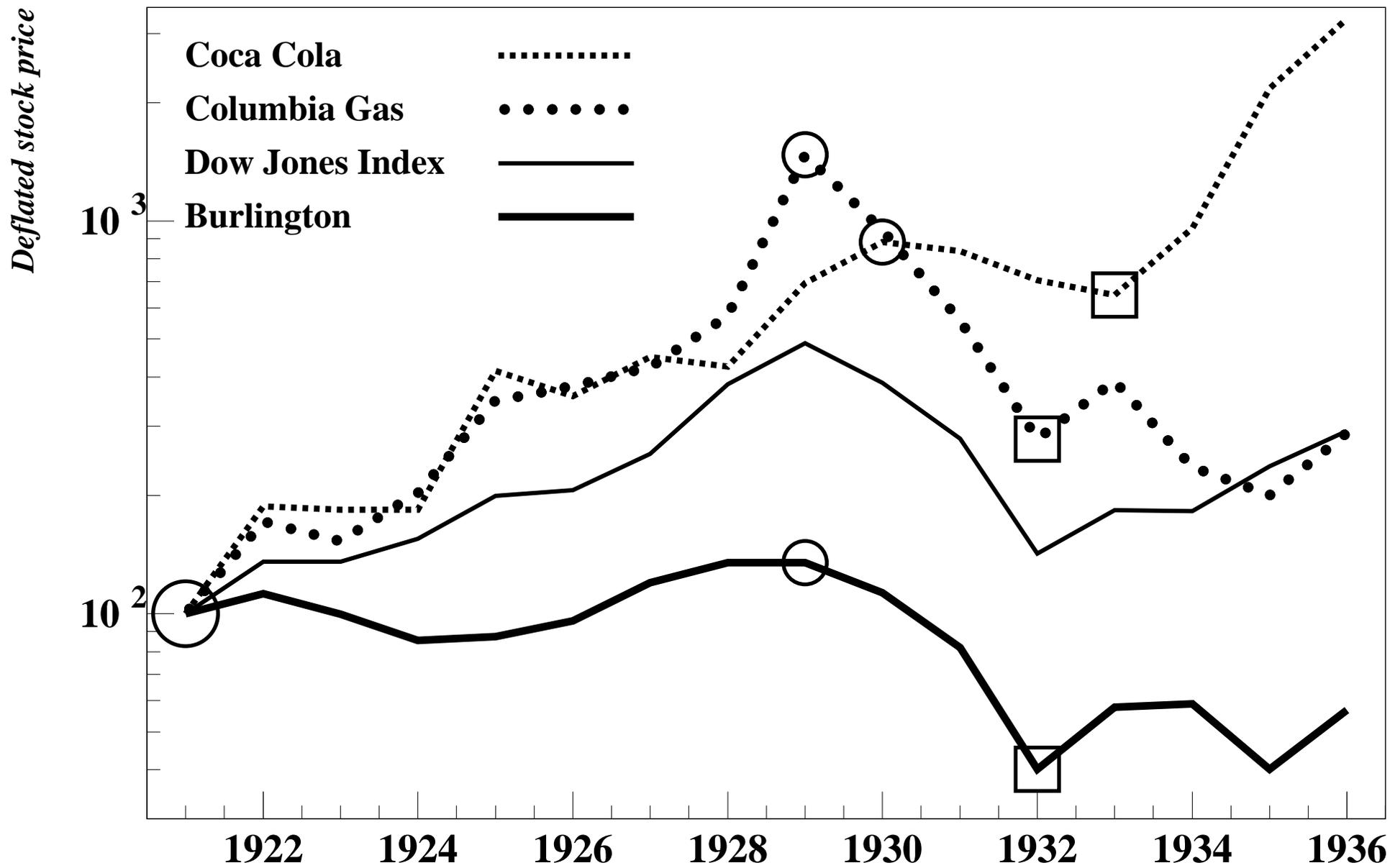
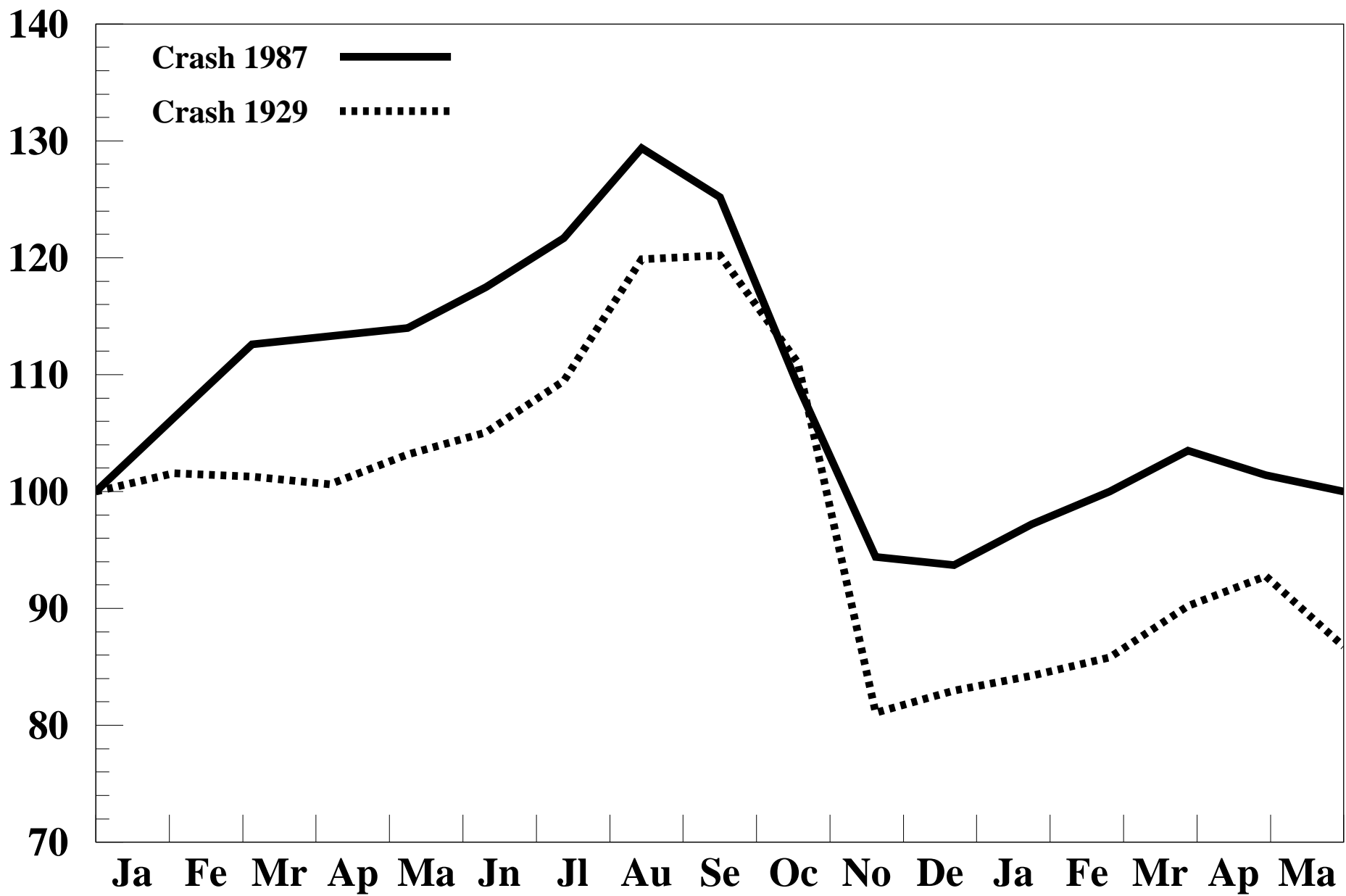
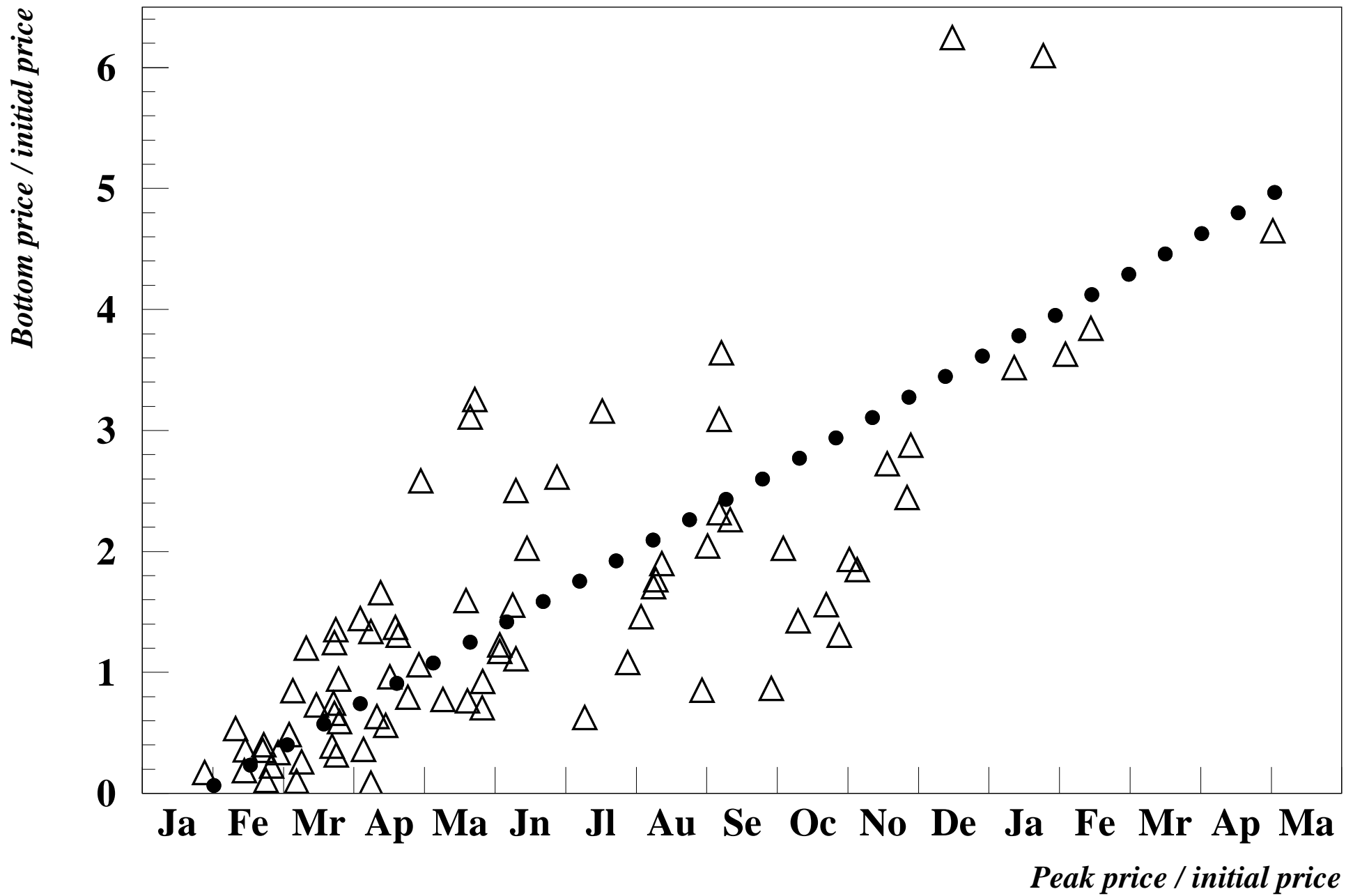
\end{document}